\documentclass{article}

\newcommand{\codid}[1]{\href{https://www.crystallography.net/cod/#1.html}{\texttt{#1}}}
\newcommand{\eg}{e.g.\@}

\usepackage{graphicx}% Include figure files
\usepackage{authblk}

\usepackage[utf8]{inputenc}
\usepackage[T1]{fontenc}
\usepackage{mathptmx}
\usepackage{etoolbox}
\usepackage[square,numbers,sort&compress]{natbib}
\usepackage{hyperref}
\usepackage{longtable}
\usepackage{float}
\hypersetup{colorlinks=true, citecolor=blue, linkcolor=blue, filecolor=magenta, urlcolor=cyan}

\usepackage{amsmath}
\let\oldAA\AA
\renewcommand{\AA}{\text{\normalfont\oldAA}}

\title{Assessing the Reliability of Foundational Machine Learning Potentials for 
Evaluating the Veracity of the Crystallography Open Database}

\author[1]{Justinas \v{S}lepavi\v{c}ius}
\author[2]{Antanas Vaitkus} 
\author[2]{Andrius Merkys}
\author[2]{Saulius Gra\v{z}ulis}
\author[1]{Linas Vil\v{c}iauskas\thanks{Email: linas.vilciauskas@ftmc.lt}}

\affil[1]{Center for Physical Sciences and Technology (FTMC), Saulėtekio al. 3, Vilnius LT-10257, Lithuania.}
\affil[2]{Institute of Biotechnology, Life Sciences Center, Vilnius University, Saulėtekio al. 7, LT-10257, Vilnius, Lithuania.}

\begin{document}

\maketitle

\begin{abstract}
The Crystallography Open Database (COD) contains over half a million experimentally determined crystal structures. Validation of crystallographic data proceeds at three hierarchical levels, the third of which assesses the physical plausibility of structures. This level has typically relied on empirical heuristics or expert judgment, which are slow and error-prone given the size of crystallographic databases and the rate at which new structures are deposited. Here we assess whether universal machine-learning interatomic potentials (MLIPs) can automate this level of validation. Of the 522 086 COD entries, 317 457 (60.8\%) passed a data-quality filter removing disordered structures, formula mismatches and unmodelled solvent. These structures were relaxed with M3GNet, and 314 153 (99.0\%) converged without a large change in cell volume. For most entries the relaxed and deposited volumes are nearly identical. Convergence and volume change thus provide simple criteria for identifying relaxations that do not confirm the deposited structure. A manual inspection of 327 structures showed that no single descriptor threshold (energy, largest force, or atomic displacement) separates valid from invalid structures. A comparison of M3GNet, CHGNet, PET-MAD and PET-OAM on 524 selected structures revealed model-specific failures: M3GNet distorts cyclopentadienyl and other $\pi$ ligands, and both M3GNet and CHGNet distort thiophene and thiazole rings, whereas the PET models preserve these motifs. All models detected missing hydrogen atoms, and at least some detected spurious hydrogens, incorrectly assigned atom types, and two previously unreported coordinate errors. No single model detected all error types. MLIP relaxation is therefore a useful but model-dependent screening tool, and combining several models is preferable to relying on any single one.
\end{abstract}

%%%%%%%%%%%%%%%%%%%%%%%%%%%%%%%%%%%%%%%%%%%%%
%%%%%%%%%%%%%%% INTRODUCTION %%%%%%%%%%%%%%%%
%%%%%%%%%%%%%%%%%%%%%%%%%%%%%%%%%%%%%%%%%%%%%

\section{\label{sec:introduction}Introduction}

The Crystallography Open Database (COD) is currently the largest comprehensive 
open-access repository containing experimental crystal structures for inorganic, 
organic, metal-organic, and mineral compounds~\cite{gravzulis2009crystallography,
gravzulis2012crystallography,gravzulis2020crystallography,Grazulis2019}.
The COD currently encompasses over half a million entries. Such a resource provides 
broad access to experimentally determined crystal structures, enables high-throughput 
data analysis and screening, facilitates the training of machine learning models, 
and offers insights into broad crystallographic and chemical trends that are not 
obvious from individual structures. Because valuable scientific insights typically 
rely on rigorous prior validation, a frequent criticism of the open data movement 
is its tendency to prioritize accessibility at the expense of data quality.
Therefore, the accumulation of insufficiently curated structural data is a 
significant problem for the usability and transferability of the COD.

The validation of crystallographic structural data fundamentally operates across 
three hierarchical levels~\cite{vaitkus2021validation}.
The first level ensures syntactic correctness, rejecting files that deviate from 
strict syntax standards. The second level governs semantic validity, ensuring 
that individual data fields conform to formal Crystallographic Information File 
(CIF) dictionary specifications and maintain logical consistency with one another.
Finally, the third level demands rigorous, domain-specific evaluation to verify 
the physical plausibility of the data. Historically, this tier has relied on 
empirical heuristics or the specialized judgment of human experts. However, given 
the scale of modern databases and the output of crystallographic research, this 
manual approach is prohibitively tedious, time-consuming, and highly susceptible 
to error. Therefore, there is a critical need to automate this final tier by 
employing atomistic models, which would allow for the rapid assessment of 
underlying crystal structures based strictly on the physical veracity of their 
atomic arrangements.

The recent emergence of foundational (or universal) machine learning interatomic
potentials (MLIPs) represents a new versatile and highly efficient tool in 
atomistic modeling~\cite{Chen2022,deng2023chgnet,batatia2025foundation,
Riebesell2025,choudhary2023unified,kovacs2023evaluation,Behler2026,
yuan2026foundation,tahmasbi2025benchmarking,yu2024systematic,marcial2026performance}.
Trained on massive datasets of first-principles calculations, these models are now 
capable of effectively describing highly diverse chemical environments across the 
entire periodic table~\cite{Riebesell2025}. Because they successfully bridge the 
accuracy of density functional theory (DFT) quantum mechanical modeling with the 
speed of empirical potentials, these models offer a highly scalable solution for 
crystallographic data curation. They could enable the automated validation of 
structural veracity across hundreds of thousands of existing and constantly 
incoming COD entries, and deliver new physical insights with relatively low 
computational overhead.

In this study, we implement this capability by deploying several popular 
foundational MLIPs such as M3GNet, CHGNet, PET-MAD and PET-OAM~\cite{Chen2022,
deng2023chgnet,MatGL,Mazitov2025,Bigi_pushing_2026} in order to evaluate their 
suitability and reliability for assessing the physical plausibility of crystal 
structures stored in the COD. By performing high-throughput structural relaxations 
across the database, we establish a computational pipeline that quantifies geometric 
and energetic parameters relevant for the third layer of crystallographic data 
validation. Finally, the presented approach allows us to systematically flag 
physically anomalous structures, providing a scalable, physics-based layer of 
verification for the crystallographic community.

%%%%%%%%%%%%%%%%%%%%%%%%%%%%%%%%%%%%%%%%%%%%%
%%%%%% MODEL AND SIMULATION METHODS %%%%%%%%%
%%%%%%%%%%%%%%%%%%%%%%%%%%%%%%%%%%%%%%%%%%%%%

\section{\label{sec:Methods}Methods}

All structures used in this study were taken from revision 297725 of the COD, 
which contained 522\,086 entries. All structures were relaxed with the 
M3GNet~\cite{Chen2022} MLIP (model M3GNet-MP-2021.2.8-PES). M3GNet has been 
trained on DFT(PBE) energies, forces and stresses from the Materials Project~\cite{Jain2013}.
The relaxations of the database were done through MatGL's~\cite{MatGL} 
interface to ASE (ase-3.25.0)~\cite{ASE} using the FIRE algorithm, with the force 
tolerance of $10^{-1}$~eV/\AA{} and 500 step limit.

Before relaxation, all entries were screened with the COD data quality filter.
The filter applies a set of rules to the information contained in each CIF and 
rejects entries that are unsuitable for relaxation with MLIPs.
These include entries with partially occupied (disordered) sites, entries whose 
declared chemical formula differs from the formula calculated from the atomic 
sites, and entries with unmodelled solvent~\cite{Sidlauskaite2026}.
The complete list of rules is given in Table~\ref{table:data_quality_rules}.
An entry may fail more than one rule in in the data quality filter.

To determine whether the deviations observed with M3GNet are specific to that 
particular model, a subset of 524 structures was additionally relaxed with 
three other universal MLIPs. The subset consists of structures selected during 
manual inspection of the M3GNet results. It primarily includes structures 
containing 
chemical motifs that M3GNet was found to describe poorly, as well as entries 
with known errors in the deposited structure model.

Each CIF of the 524 structure subset was expanded to the full content of the 
unit cell in space group P$1$ with the \texttt{cif\_fill\_cell} program of 
\texttt{cod-tools} version~3.12.0~\cite{Vaitkus2021}.
This program applies the symmetry operators given in the CIF to the atoms of 
the asymmetric unit. The Cartesian atomic coordinates obtained from the 
expanded cells were used as input for the relaxations using the LAMMPS suite of 
programs~\cite{LAMMPS}. M3GNet and CHGNet~\cite{deng2023chgnet} (model 
CHGNet-MPtrj-2024.2.13-11M-PES), another MLIP from the MatGL library, were 
run through AdvanceSoft Corp.\ interface.~\cite{ASC-LAMMPS} PET-MAD
~\cite{Mazitov2025} (PET-MAD-S-1.5.0) and PET-OAM~\cite{Bigi_pushing_2026} 
(PET-OAM-XL-1.0.0) were run through the \texttt{metatomic} interface to 
LAMMPS~\cite{bigi_metatensor_2026,Metatomic-LAMMPS}. LAMMPS was the preferred 
relaxation engine throughout this work, but no LAMMPS interface to MatGL models 
was available at the start of this project, so ASE was used 
instead. By the time the detailed comparison of the selected subset was performed
a LAMMPS interface to MatGL had become available, and LAMMPS was used for all 
four models in this comparison.

Following the cell relaxation example distributed with LAMMPS~\cite{LAMMPS-min-box}, 
each structure was relaxed in four consecutive stages, each adding additional 
degrees of freedom: (i) relaxation of the atomic positions in a fixed cell 
(\texttt{minimize} without additional fixes); (ii) relaxation of the atomic 
positions together with isotropic scaling of the cell (\texttt{fix box/relax iso}), 
which drives the hydrostatic pressure, i.e.\ one third of the negative trace of the 
stress tensor, to the target value; (iii) relaxation with independent scaling of the 
three cell lengths at fixed cell angles (\texttt{fix box/relax aniso}), which drives 
the three normal components of the stress tensor to the target value; (iv) relaxation 
of the full cell, including the cell angles (\texttt{fix box/relax tri}), which 
drives all six independent components of the stress tensor to the target value.
In all stages the target stress was zero. Each stage used the \texttt{cg} minimizer 
with an energy tolerance of 0, a force tolerance of $10^{-8}$~eV/\AA{}.
The \texttt{box/relax} parameters were set to \texttt{vmax}~=~0.0001 and 
\texttt{nreset}~=~0. The structure was recorded before and after each stage.
The staged procedure serves two purposes. First, it makes it possible to attribute 
the structural changes to the relaxation of the atomic positions, the cell volume 
or the cell shape. Second, it makes the minimization procedure more robust, because 
the cell degrees of freedom are introduced only after the atomic forces have been 
largely relaxed. After stage (ii) the cell is at zero pressure but may 
still carry non-zero normal 
and shear stresses, and after stage (iii) the shear stresses remain. Only the 
structures obtained after stage (iv) can be compared as relaxed structures.
The force tolerance in LAMMPS applies only to the atomic forces. All four 
models were applied with the same staged procedure and the same 
minimization settings. Differences between the relaxed structures can therefore be 
attributed to the MLIP models rather than to the relaxation protocol.

%%%%%%%%%%%%%%%%%%%%%%%%%%%%%
%%%%%%%% RESULTS %%%%%%%%%%%%
%%%%%%%%%%%%%%%%%%%%%%%%%%%%%

\section{\label{sec:Results}Results and Discussion}

\subsection{Pre-selection}

\begin{figure}[htbp]
    \includegraphics[width=\linewidth]
{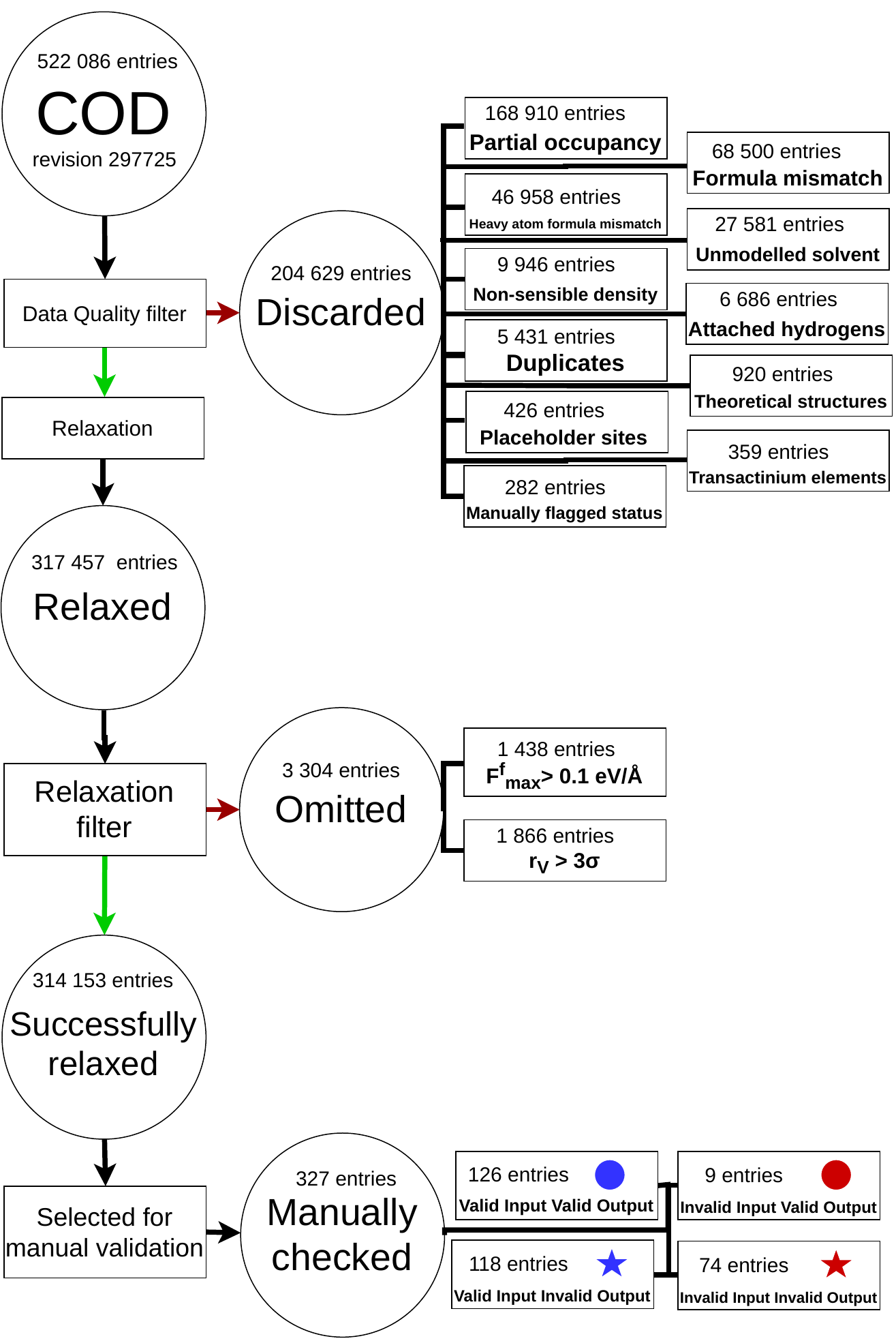}
    \caption{The workflow used in the benchmarking of M3GNet and filtering 
    of candidate structures. A full explanation of the data quality filter 
    rules is presented in Table~\ref{table:data_quality_rules}. Each entry 
    may fail more than one selection rule.}
    \label{fig:schema}
\end{figure}

The full workflow is presented in Figure~\ref{fig:schema}.
A total of 317\,457 (60.8\%) COD entries passed the data quality filter and 
were used as inputs for structural relaxation using the MLIP.
168\,910 of the rejected entries are disordered, i.e.\ they contain 
site occupancies below or exceeding 1.
Relaxing a disordered structure requires constructing one or more 
ordered models, for example supercells with explicit site occupations or 
quasi-random structures, which was outside the scope of the present work.
The second most frequent reason for rejection was a mismatch between the 
declared chemical formula and the formula calculated from the atomic sites, 
which affected 68\,500 entries.
In 21\,614 of these, only the number of hydrogen atoms differed.
Hydrogen atoms scatter X-rays weakly, so their positions are difficult to 
determine from X-ray diffraction data and are not always included in the 
deposited structure model.
The remaining 46\,958 entries also differ in the numbers of non-hydrogen 
atoms, which indicates a more serious inconsistency between the declared 
formula and the list of atomic sites.
Both kinds of mismatch excluded an entry from relaxation, because a 
structure that has incorrectly labelled, missing or additional atoms, 
including hydrogen, cannot completely represent the deposited 
compound.
A total of 27\,581 entries contain unmodelled solvent.
In these entries, solvent molecules present in the crystal are absent 
from the structure model, so the structure as deposited is incomplete 
and cannot be relaxed meaningfully using an atomistic representation.
Table~\ref{table:data_quality_rules} lists all rules of the data 
quality filter.

\begin{longtable}{lrrp{0.42\textwidth}}
\caption{\label{table:data_quality_rules} Rules of the data quality 
filter.
For each rule, the table gives the number of entries rejected by it 
and this number as a percentage of all rejected entries.
An entry may be rejected by more than one rule, so the counts sum to 
more than the number of rejected entries and the percentages to more 
than 100\%.}\\
\hline
Rule & Entries & Percentage & Explanation \\
\hline
\hline
\endfirsthead
\multicolumn{4}{c}{\tablename\ \thetable{} -- continued from previous page}\\
\hline
Rule & Entries & Percentage & Explanation \\
\hline
\hline
\endhead
\hline
\multicolumn{4}{r}{continued on next page}\\
\endfoot
\hline
\endlastfoot
Partial occupancy                    & 168\,910 & 82.5\% & Disordered structures with site occupancies below or exceeding 1.\\
\hline
Formula mismatch                     &  68\,500 & 33.5\% & The presented and calculated formulas are different, indicating an issue with the CIF.\\
\hline
Heavy atom formula mismatch          &  46\,958 & 22.9\% & The presented and calculated formulas are different even when hydrogen atoms are excluded.\\
\hline
Unmodelled solvent                   &  27\,581 & 13.5\% & The crystal structure reported does not contain solvent molecules, where they should be. This makes the structure unphysical and, hence, not viable for confirmation {\it via} relaxation. \\
\hline
Non-sensible density                 & 9\,946   & 4.9\%  & Structure's density obtained from the reported volume and atom mass is lower than 0.17 $\text{g/cm}^3$ (solid Helium) or higher than 23 $\text{g/cm}^3$ (Osmium).\\
\hline
Attached hydrogens                   &   6\,686 &  3.3\% & The structure contains the \texttt{\_atom\_site\_attached\_hydrogens} flag and a non-zero value, indicating that some atoms contain a hydrogen atom the coordinates of which are not provided.\\
\hline
Duplicates                           &   5\,431 &  2.7\% & Structures that are duplicates of another structure already present in the COD and used in further validation.\\
\hline
Theoretical structures               &     920  &  0.4\% & Structures that are obtained theoretically or by computation and not experimentally verified.\\
\hline
Placeholder sites                    &     426  &  0.2\% & The structures that contain placeholder or `dummy' sites.\\
\hline
Unsupported elements               &     359  &  0.2\% & The structure contains an element beyond M3GNet's training 
data (Z > 94).\\
\hline
Manually flagged status              &     282  &  0.1\% & Structures with a non-default COD curator status 
(retracted, warnings, or errors)\\
\hline
\hline
\textbf{Total} & 204\,629 & & \\
\hline
\end{longtable}

\subsection{Filtering of relaxed structures}

To keep the relaxations short, we used a coarse convergence 
criterion of the maximum final force, specifically 
$F^{\mathrm{f}}_{\max}<0.1$~eV/\AA{}.
A relaxation that fails to converge suggests that the structure 
is far from a minimum of the M3GNet potential energy surface or 
that it contains atomic configurations that are poorly represented 
in the training data of the model.
Of the 1\,438 unconverged relaxations, 752 (52\%) ended with 
$F^{\mathrm{f}}_{\max}\leq0.2$~eV/\AA{} and 1\,148 (79.8\%) with 
$F^{\mathrm{f}}_{\max}\leq0.5$~eV/\AA{}.
For most of the unconverged structures the minimization therefore 
approached a minimum but did not meet the criterion.
Because the criterion was coarse, 1\,019 structures were not changed 
at all: their $F^{\mathrm{i}}_{\max}$ was already below 0.1~eV/\AA{}.
These deposited structures are already close to a minimum of the M3GNet 
potential energy surface. 
Examples of structures that did not change during the relaxation 
include 24 different structures of $\alpha$-iron, 20 diamond structures, 
many rock-salt oxides and halides (MgO, CaO, NaCl, NaBr, NaF), elemental 
and binary metallic structures (Al$_4$, Cu$_4$, NiTi, YbAu, LuMg, ErCu) 
and a perovskite (TiPbO$_3$).

\begin{figure}[H]
    \includegraphics[width=\linewidth]
{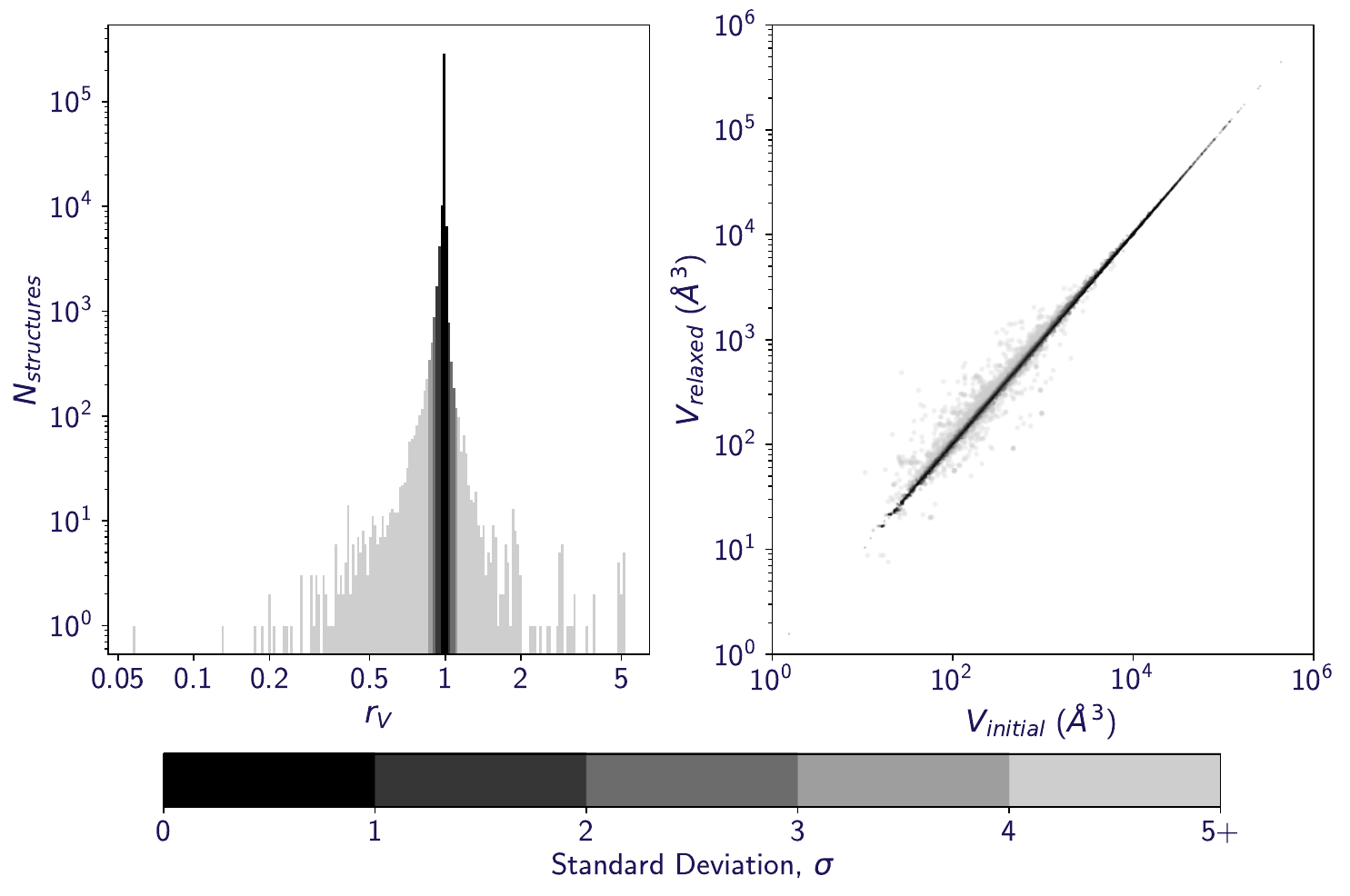}
    \caption{Comparison of the cell volumes before and after 
    relaxation. Left: histogram of the volume ratio $r_V$ 
    (Equation~\ref{eq:volratio}).
Right: relaxed versus initial cell volume on logarithmic axes.
On these axes, $\log r_V$ is proportional to the distance of a 
point from the diagonal $y=x$, so the histogram corresponds to 
a projection of the scatter plot onto the direction 
perpendicular to the diagonal.}
    \label{fig:VolumeCovariance}
\end{figure}

The second criterion is based on the change in cell volume upon 
relaxation, which summarizes the change of the lattice in a single 
number, although it does not capture changes in cell shape at 
constant volume.
A large volume change shows that the deposited structure is far 
from a minimum of the M3GNet potential energy surface.
This may indicate an error in the deposited structure, such as 
missing atoms or unmodelled solvent, but it may also reflect a 
limitation of the model: M3GNet was trained on PBE data, which 
systematically overestimate lattice parameters and do not include 
dispersion interactions.
In addition, the relaxation corresponds to 0~K, whereas most 
structures were determined at finite temperature.
We characterize the volume change by the ratio
\begin{equation}
r_V = \frac{V_\mathrm{relaxed}}{V_\mathrm{initial}},
  \label{eq:volratio}
\end{equation}

\noindent where $V_\mathrm{initial}$ is the cell volume given in 
the CIF and $V_\mathrm{relaxed}$ is the volume after relaxation.
The mean $\mu$ and the standard deviation $\sigma$ of $r_V$ were 
computed directly from the values for all relaxed structures.
The distribution of $r_V$ is sharply peaked at $r_V=1$ 
(Figure~\ref{fig:VolumeCovariance}), so for most structures the 
relaxed and deposited cell volumes are nearly identical.
It is consistent with the PBE data used to train M3GNet~\cite{Pohls2020}.
For most COD entries, M3GNet reproduces the experimental cell 
volume very well, across a chemically diverse set of structures.
Structures with $|r_V-\mu|>3\sigma$, corresponding to a volume 
change of $r_V<0.88$ and $r_V>1.15$, were marked as problematic 
and did not pass the filter for further analysis.
For the retained structures, it means that the volume changes 
arise from small adjustments of the interatomic distances rather 
than from a more global rearrangement of the structure.

The entries discarded by the volume criterion are dominated by 
structures measured away from ambient conditions. Expansions 
(1,468 of 1,866 entries) are led by structures measured under 
high pressure, for example periclase (MgO), relaxed from 
a single-crystal compression study reaching 118~GPa, as well as 
several high-pressure silicate and germanate phases. 
Contractions (398 of 1,866) are dominated by iron-rich structures 
measured at high temperature, such as the face-centred-cubic 
$\gamma$-iron phase, stable only above 912\textdegree{}C. 
Because M3GNet was trained on static 0 K calculations at zero 
pressure, and no external pressure was used in relexation,
the procedures relaxed such entries to a different volume 
regardless of whether the deposited structure is correct.
After both criteria were applied, 314\,153 
of the 317\,457 relaxed structures (99.0\%) were retained and 
we refer to these as structures that passed a successful 
relaxation.

Relaxation alone cannot establish whether a deposited 
structure is correct, and interpreting its outcome requires 
distinguishing two properties of each entry: whether the 
deposited structure is a valid model of the compound, and 
whether relaxation changes it substantially.
Their combinations give four outcomes, analogous to the 
cells of a confusion matrix: 
(i) a valid structure that is not changed substantially, 
which reproduces the deposited structure, corresponding to 
the true negative of the confusion matrix; 
(ii) a valid structure that is changed substantially, which 
is a false positive caused by a limitation of the model; 
(iii) an invalid structure that is changed substantially, 
representing a detected error, a true positive; and 
(iv) an invalid structure that is not changed substantially, 
which is a missed error - a false negative.

False positives (ii) arise for two reasons.
First, MLIPs are trained on DFT energies, forces and stresses 
and therefore reproduce DFT rather than experimental geometries.
PBE structures themselves deviate from experimental ones, so 
some difference between the deposited and the relaxed structures 
is expected even for correct entries.
Second, chemical motifs that are underrepresented in the training 
data may be described poorly by the model.
In this work, the clearest example is the inability of M3GNet 
to preserve the structure of metallocenes, in which, for example, 
a metal atom is sandwiched between two cyclopentadienyl (Cp$^-$) 
ligands.
Upon relaxation with M3GNet, the C--C bonds of the Cp$^-$ rings 
are elongated and the hydrogen atoms are displaced out of the 
ring plane (Figure~\ref{fig:7119083-m3gnet-cp-distortion}).
Such model-specific failures are discussed in 
Section~\ref{subsec:M3gNet-fail}.
A valid structure that is distorted by an MLIP is therefore 
neither confirmed nor shown to be incorrect.

Some invalid entries that were not removed by the data 
quality filter nevertheless converged and had little volume 
change, passing both filtering criteria.
If such an entry is not changed substantially by relaxation 
(iv), the relaxation provides no evidence either way, and 
the error must be found by other means.
In contrast, when an invalid structure relaxes to a more 
physical one (iii), the change is a successful detection 
of the problematic entry.
However, this should be taken with caution: the relaxed 
structure is still not an experimentally determined model 
of the compound.

We also attempted to construct a further criterion from 
the quantities recorded during relaxation: the initial 
and final energies per atom ($E^{\mathrm{i}}$ and 
$E^{\mathrm{f}}$), the initial and final maximum atomic 
forces ($F^{\mathrm{i}}_{\max}$ and $F^{\mathrm{f}}_{\max}$), 
the root-mean-square displacement of all atoms (RMSD) and the 
maximum displacement of a single atom ($d_{\max}$).
Because the energy per atom ranged from $-10$ to $10^{12}$~eV/atom, 
it is displayed using the symmetric logarithm

\begin{equation}
\operatorname{symlog}_{10}(E)=\operatorname{sign}(E)\,\log_{10}\!
\left(1+|E|\right),
  \label{eq:symlog}
\end{equation}

\noindent which preserves the sign and compresses the range 
for better visualization.
Energies of the order of $10^{12}$~eV/atom are clearly 
unphysical and arise from atoms that are too close to 
each other in the input structure.

Unlike convergence and volume change, these quantities did 
not provide a straightforward criterion.
To relate them to the validity of the structures, we 
manually inspected 327 structures, selected to cover as 
much of the six-dimensional space of the descriptors as 
possible.
Each structure received two labels: whether the deposited 
(input) and relaxed (output) structures are chemically and 
crystallographically valid.
In terms of the four outcomes defined above, a valid input 
with an invalid output corresponds to a false positive and 
an invalid input to an error that relaxation should detect.

\begin{figure}[H]
    \includegraphics[width=\linewidth]{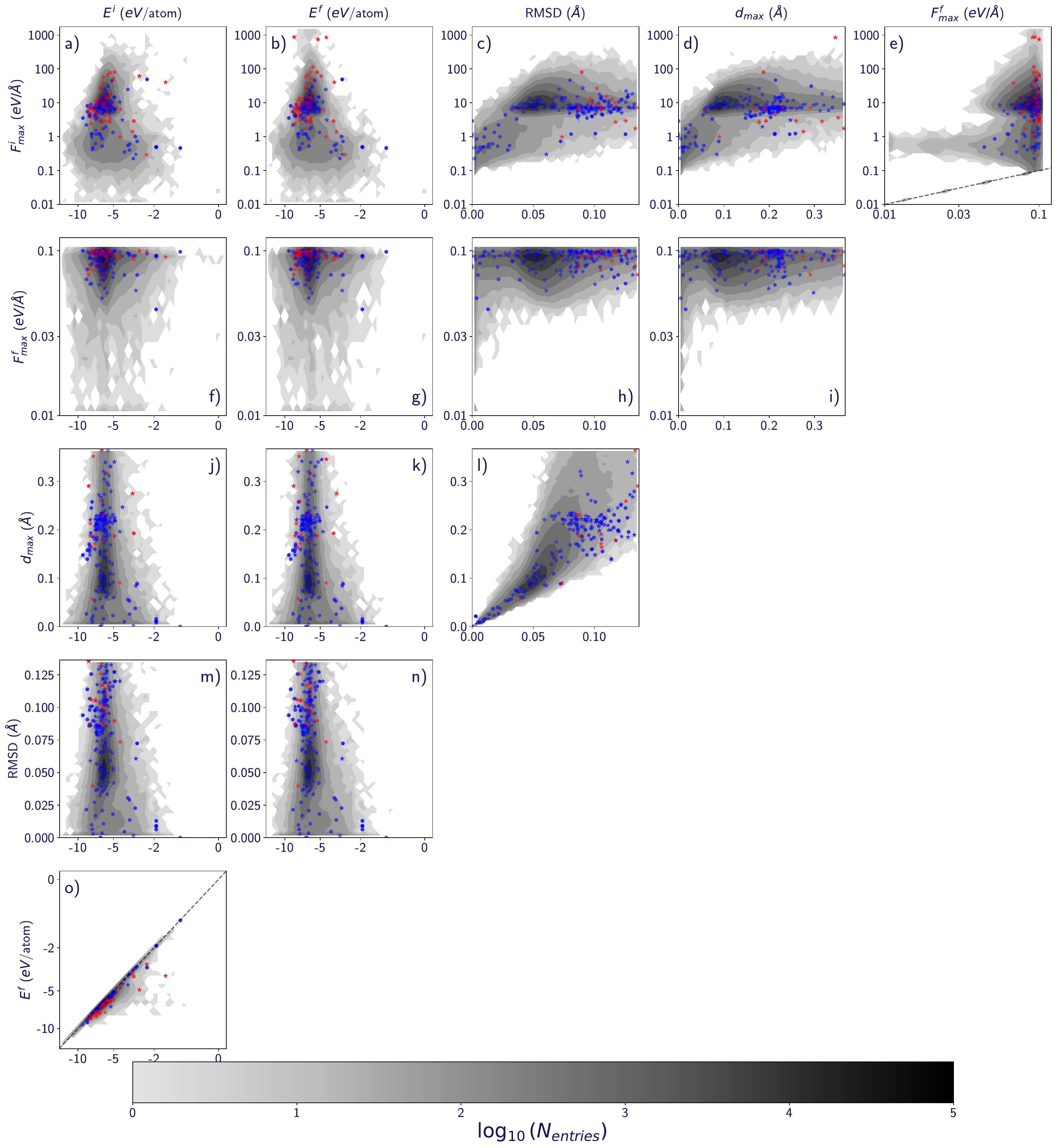}
    \caption{Pairwise distributions of the relaxation descriptors
$E^{\mathrm{i}}$, $E^{\mathrm{f}}$, $F^{\mathrm{i}}_{\max}$, 
$F^{\mathrm{f}}_{\max}$, RMSD and $d_{\max}$ for all successfully 
relaxed structures.
Symbols mark the manually inspected structures: the colour 
indicates the validity of the input structure (blue, valid; 
red, invalid) and the shape the validity of the relaxed structure 
(circle, valid; star, invalid).
In the panels comparing initial with final values of $F_{\max}$ 
and of $E$, the dashed line is the diagonal $y=x$.
Only the most populated region is shown; the full ranges are given 
in Figure~\ref{sup-fig:All_Graphs_full}.}
    \label{fig:All_Graphs_mini}
\end{figure}

The pairwise distributions of the descriptors are shown 
in Figure~\ref{fig:All_Graphs_mini}.
For all structures, both $E$ and $F_{\max}$ 
decreased or remained unchanged during relaxation, as 
expected (Figure~\ref{fig:All_Graphs_mini}o and 
\ref{fig:All_Graphs_mini}e).
Most structures lie close to the diagonal in 
Figure~\ref{fig:All_Graphs_mini}o, i.e.\ their energy 
changed little, whereas in Figure~\ref{fig:All_Graphs_mini}e 
most structures lie far from the diagonal, 
with $F^{\mathrm{i}}_{\max}$ about 1000 times greater 
than $F^{\mathrm{f}}_{\max}$.

These descriptors weakly separate the valid and invalid 
manually inspected structures.
$F^{\mathrm{f}}_{\max}$ does not separate them at all, 
which is expected: the minimizer stops at the first step 
below the convergence criterion, so nearly all converged 
relaxations end with $F^{\mathrm{f}}_{\max}$ just below 0.1~eV/\AA{}.
For $F^{\mathrm{i}}_{\max}$, the most informative threshold 
is 50~eV/\AA{}: 33 of 34 inspected structures with 
$F^{\mathrm{i}}_{\max}>50$~eV/\AA{} have an invalid input 
or output whereas the inspected structures below the threshold 
are a mixture of valid and invalid ones.
Forces of this magnitude are far larger than those expected 
in a physically reasonable structure and indicate atoms that 
are too close to each other, so the threshold also has 
a physical interpretation.
It is nevertheless incomplete: many invalid structures lie 
below it, most of them with a valid input and an invalid output, 
i.e.\ structures that M3GNet does not describe correctly.
Moreover, the relaxations of structures above the threshold still 
converged to $F^{\mathrm{f}}_{\max}<0.1$~eV/\AA{}, so convergence 
alone does not imply that the relaxed structure is valid.

A similarly weak threshold applies to the energies, $E^i>0$~eV/atom.
Compared with the threshold on $F^{\mathrm{i}}_{\max}$, this 
catches 6 fewer structures, and all $E^i$ filtered structures 
are already caught by the $F^{\mathrm{i}}_{\max}$ threshold.
In the planes of $F$ or $E$ against RMSD or $d_{\max}$, the two 
groups of quantities show no mutual dependence, so any threshold 
in these planes reduces to a threshold on a single quantity.
In the RMSD--$d_{\max}$ plane, the valid and invalid structures 
are not separated by a straight line either, which suggests 
that a nonlinear classifier would be required.

\subsection{Comparison with CHGNet and PET models}
\label{sec:mlip-comparison}

Manual inspection of the M3GNet relaxations revealed systematic 
distortions of several chemical motifs, of which the metallocenes 
are the most prominent example.
To determine whether such deviations are specific to M3GNet, and 
how the choice of model affects the detection of errors in the 
deposited structures, we relaxed a subset of 524 structures with 
three other MLIPs: CHGNet~\cite{deng2023chgnet}, a more recent 
model from the MatGL library, and the PET models PET-MAD and 
PET-OAM~\cite{Mazitov2025}.
The subset comprises three groups of structures: structures that 
were inspected manually, structures containing chemical motifs 
identified as challenging for M3GNet, and structures with known 
errors in the deposited structure model.

\subsubsection{Model cross-comparison}
\label{subsec:crosscomparisons}

The most direct comparison between the M3GNet, CHGNet, PET-MAD 
and PET-OAM models is that of the relaxed cell volumes with 
the deposited volume (Figure~\ref{sup-fig:Multi_Volume}).
All models agree closely on the cell volume.
The closest agreement is among the deposited volume and the 
volumes obtained with M3GNet and CHGNet: these three pairs 
deviate very little from the diagonal and have concordance 
correlation coefficients (CCC)~\cite{Lin1989} close to unity.
The CCC measures agreement with the line $y=x$, unlike the 
Pearson coefficient, which measures agreement with any 
straight line.
The PET models deviate more from the diagonal, both from each 
other and from the deposited volume.
Close agreement with the deposited volume is partly expected, 
because the structures selected for the comparison were drawn 
from those whose volume did not change substantially in the 
M3GNet relaxation of the full database.

\begin{figure}[H]
    \includegraphics[width=0.9\linewidth]{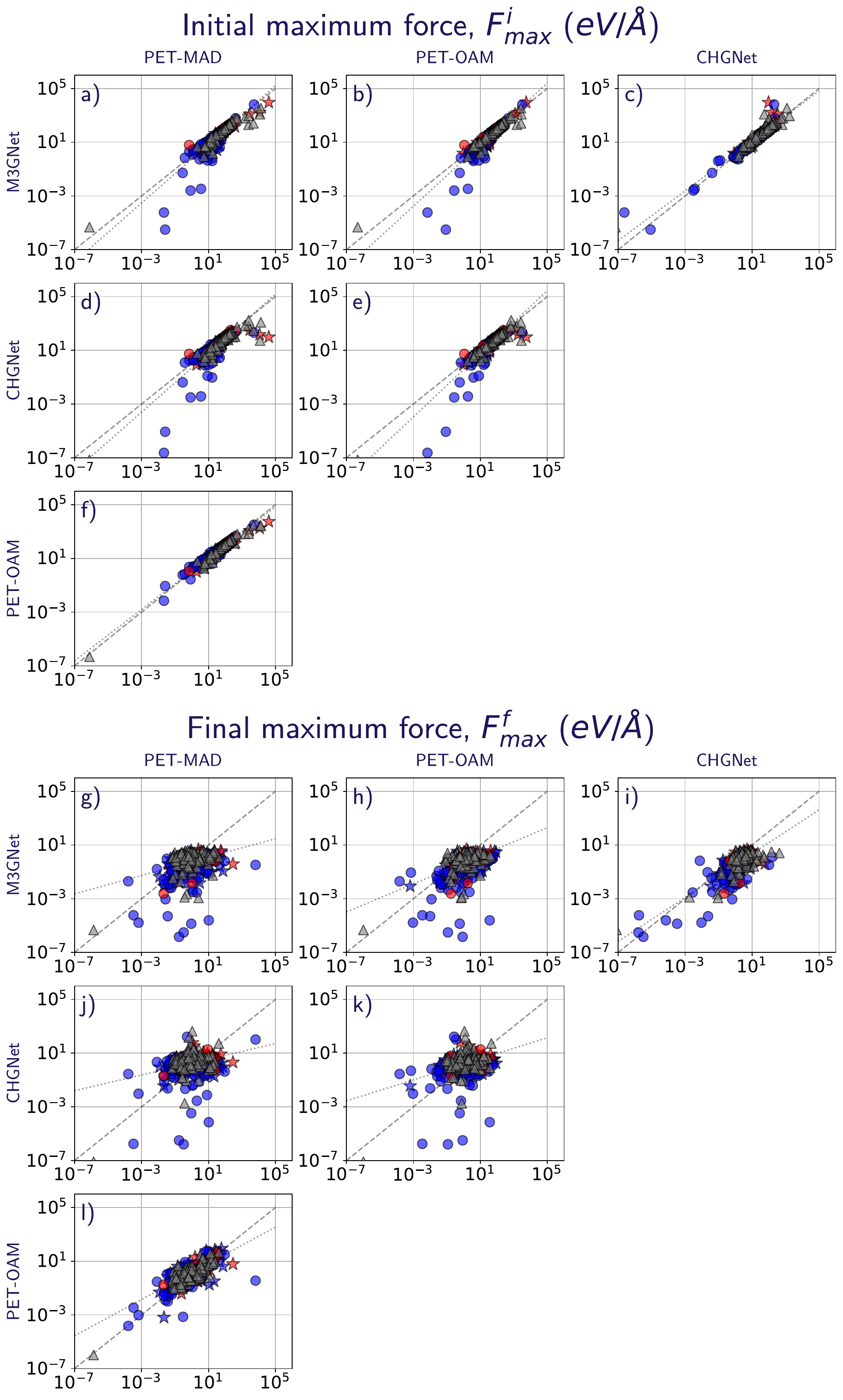}
    \caption{Pairwise correlation of initial and final 
    largest atomic forces,
$F^{\mathrm{i}}_{\max}$ and $F^{\mathrm{f}}_{\max}$, obtained 
with the four MLIPs.
For the inspected structures, the colour of the symbol indicates 
the validity of the input structure (blue, valid; red, invalid) 
and its shape the validity of the relaxed structure (circle, 
valid; star, invalid).
The dashed line is the diagonal $y=x$ and the dotted line is the 
linear fit to $y=\log_{10}F_{\max}$ (vertical axis) against 
$x=\log_{10}F_{\max}$ (horizontal axis); the fitted equation and 
the coefficient of determination $R^2$ are presented in the 
Supplementary Information Table~\ref{sup-table:multi_force_stats}.}
    \label{fig:Multi_Force}
\end{figure}

Because there is no reference value for the forces, the largest 
atomic forces can only be compared between models 
(Figure~\ref{fig:Multi_Force}).
For each pair of models we fitted a straight line to the 
$\log_{10}$ values and report its slope $m$ and the coefficient 
of determination $R^2$ of the fit, which gives the fraction of 
the variance of one model values explained by the linear 
relation to the other.
For $F^{\mathrm{i}}_{\max}$, the models agree well for 
most structures, with $R^2>0.75$ for every pair.
The agreement is closest within the two model families, 
with $R^2=0.94$ for the MatGL models and $R^2=0.96$ for the 
PET models.
General agreement is expected, since all four models were 
trained on DFT data, although the training sets differ in 
composition and in the exchange--correlation functional.

A large number of structures have a lower 
$F^{\mathrm{i}}_{\max}$ with the MatGL models than with 
the PET models.
All of them lie at the low end of the range, with 
$F^{\mathrm{i}}_{\max}<1$~eV/\AA{} for CHGNet and M3GNet.
This may be related to the differences in the cut-off radii 
$r_c$ of the models, which are 5~\AA{} for CHGNet and M3GNet, 
8~\AA{} for PET-MAD and 10~\AA{} for PET-OAM~\cite{Bigi_pushing_2026}.
However, PET-MAD and PET-OAM agree with each other in this 
range although their cut-off radii also differ, so the cut-off 
radius alone is unlikely to explain the difference.

\begin{figure}[H]
    \centering
    \includegraphics[width=0.7\linewidth]{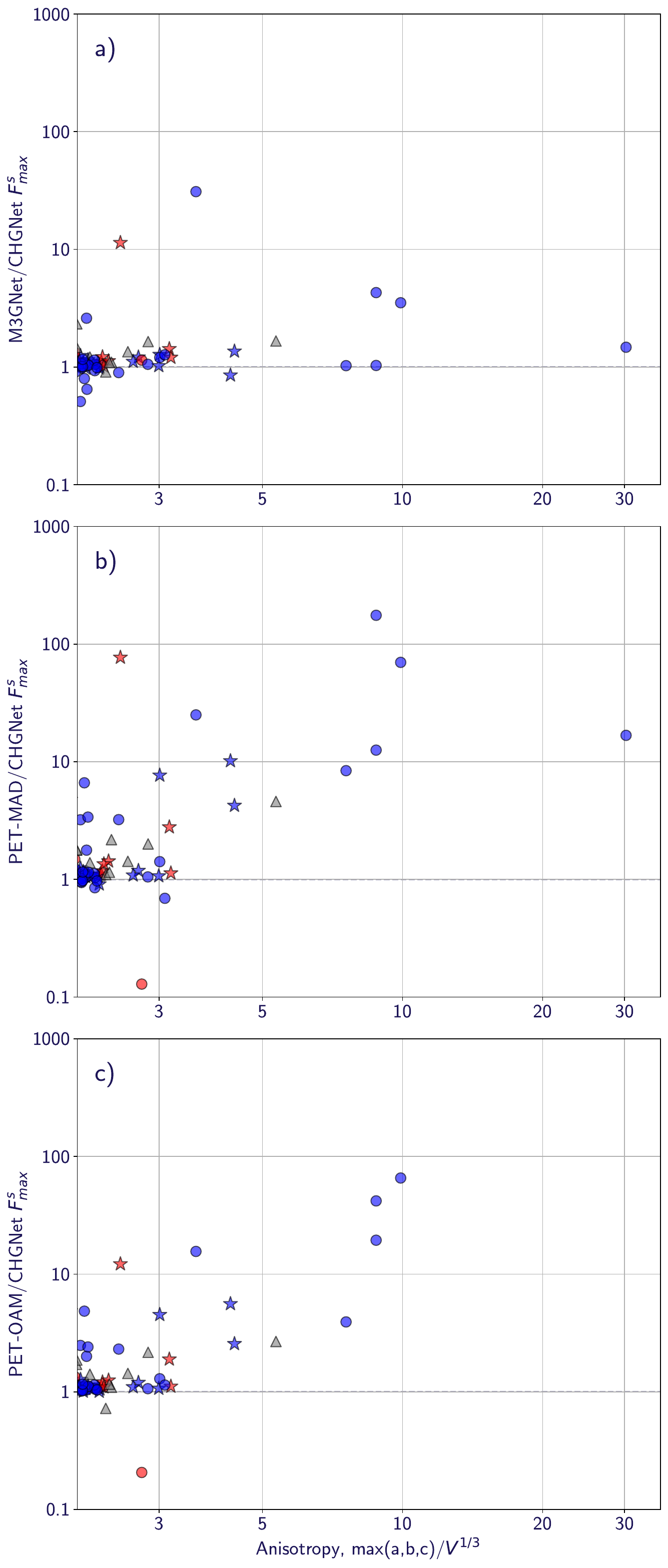}
    \caption{Ratio of the $F^{\mathrm{i}}_{\max}$ obtained 
    with CHGNet
to that obtained with each of the other models, as a function 
of the lattice anisotropy.
Symbols are the same as in Figure~\ref{fig:Multi_Force}.}
    \label{fig:Anisotropy}
\end{figure}

A second difference concerns fewer structures: for 
$F^{\mathrm{i}}_{\max}\geq10$~eV/\AA{}, CHGNet consistently 
predicts smaller forces than the other models, particularly 
the PET models.
This difference correlates with the anisotropy of the lattice, 
defined here as $\max(a,b,c)/V^{\frac{1}{3}}$ (Figure~\ref{fig:Anisotropy}).
For all structures with an anisotropy of 4 or more, CHGNet 
predicts a lower $F^{\mathrm{i}}_{\max}$ than the PET models, 
and in most cases a lower one than M3GNet.

The values of $F^{\mathrm{f}}_{\max}$ are much less correlated.
Most structures relax to $10^{-3}~\mathrm{eV/\AA}<F^{\mathrm{f}}_{\max}<10^{2}~\mathrm{eV/\AA}$, with no clear relation between the models.
The agreement is again best within the model families, but 
even there it is weak.
Weak agreement is expected, because the final forces depend 
mainly on where the minimizer stops rather than on the model.

The energies show a pattern similar to $F^{\mathrm{i}}_{\max}$, 
but with larger scatter, so the agreement between the models 
is poorer.
In particular, the lower values of CHGNet for anisotropic lattices 
are more pronounced in $E^{\mathrm{i}}$ than in 
$F^{\mathrm{i}}_{\max}$ (Figure~\ref{sup-fig:Multi_Energy}).
The values of $E^{\mathrm{f}}$, in contrast, lie close to the diagonal.
The main exception is a single structure for which the final 
PET-MAD energy is orders of magnitude higher than that obtained 
with the other models.

The displacements, RMSD and $d_{\max}$ (Figure~\ref{sup-fig:Multi_Distance}), 
reveal a further peculiarity of CHGNet: it displaces the atoms 
considerably less than the other models.
The slopes $m$ of the fits of M3GNet against the PET models lie 
between 0.84 and 1.2, close to the diagonal, whereas all fits 
involving CHGNet have $m>2.2$ or $m<0.45$, much further away 
from the diagonal, most markedly against the PET models.
Despite these well-defined slopes, the scatter about the fits 
is large, with $R^2\approx0.6$ for RMSD and $0.55<R^2<0.67$ 
for $d_{\max}$.
The exception is the pair of PET models, with $R^2\approx0.85$ 
for both quantities.
The smaller displacements obtained with CHGNet are consistent 
with a softer potential energy surface, which yields smaller 
forces for a given distortion, so that the convergence criterion 
is reached closer to the starting geometry.
Systematic softening has been reported for universal MLIPs 
trained on Materials Project data~\cite{Deng2025}.

\subsubsection{Analysis of problematic M3GNet cases}
\label{subsec:M3gNet-fail}

Manual inspection of the M3GNet relaxations revealed a 
systematic misrepresentation of structures containing certain 
delocalised bonds.
The two most frequently encountered distorted motifs were 
examined in more detail to determine whether the distortions 
are specific to M3GNet or are also produced by the other models.

\begin{figure}[H]
\centering
    \includegraphics[
width=\linewidth, trim={0 11cm 0 11cm}, clip ] {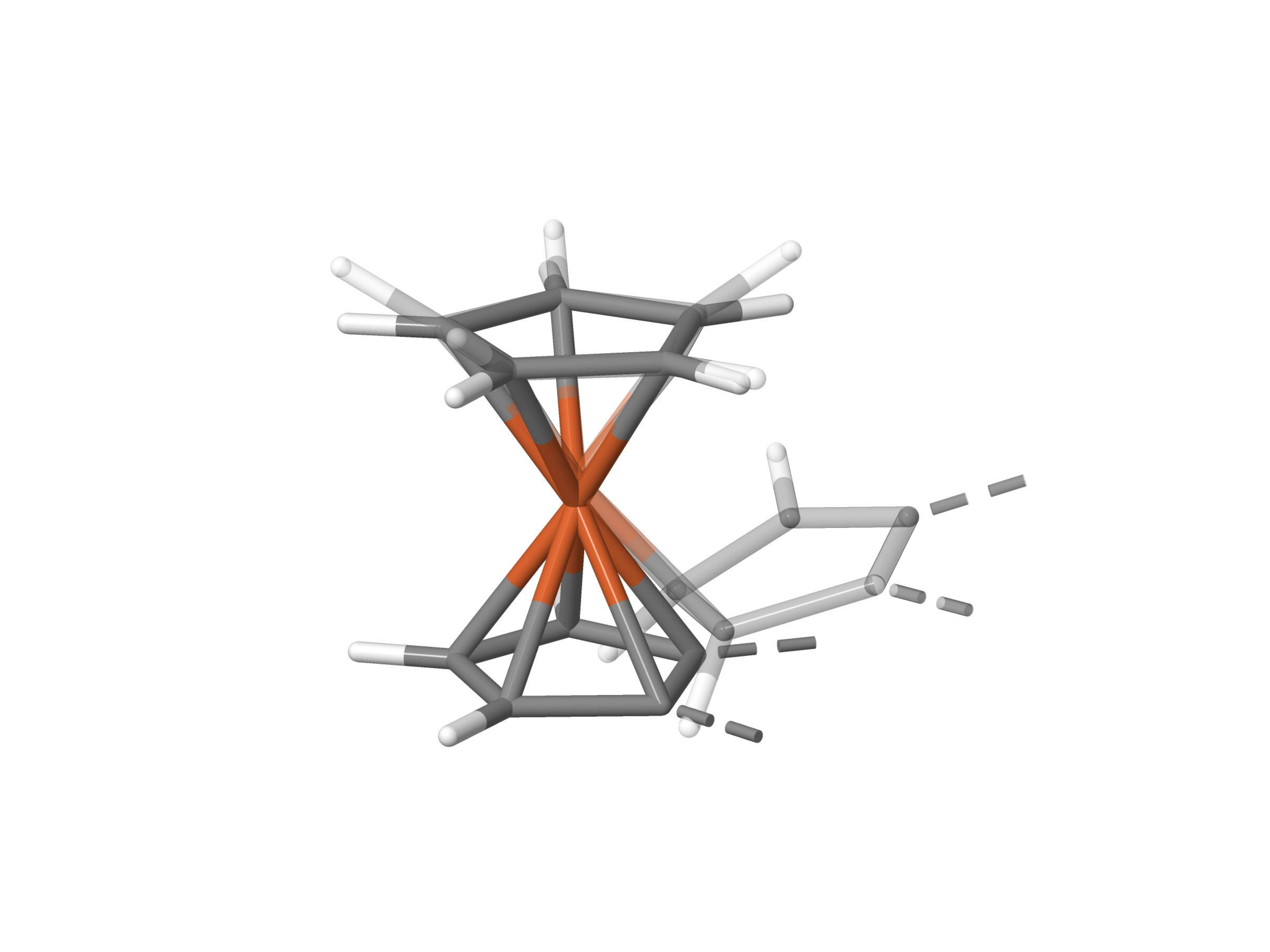}
    \caption{
Typical distortion of the cyclopentadienyl ($\text{Cp}^-$) 
ligand after relaxation with M3GNet, as observed in COD entry \codid{7119083}.
The deposited (opaque) and relaxed (translucent) structures 
are aligned by minimizing the RMSD between the five $\text{Cp}^-$ carbon atoms.
Dashed lines indicate bonds to atoms omitted from the 
illustration for clarity.
}
    \label{fig:7119083-m3gnet-cp-distortion}
\end{figure}

The first motif comprises $\pi$ ligands in metal coordination 
complexes, specifically metallocenes.
After relaxation with M3GNet, cyclopentadienyl ligands were 
consistently distorted: the hydrogen atoms were displaced out 
of the plane of the carbon atoms, and the C--C bonds were 
elongated to lengths characteristic of 
\mbox{\textit{sp}$^2$--\textit{sp}$^2$} 
single bonds rather than of aromatic bonds 
(Figure~\ref{fig:7119083-m3gnet-cp-distortion}).
A similar bond elongation was observed in other $\pi$ ligands, 
such as dialkyl(butadiene) (\eg{} COD entry \codid{4077522}).
These distortions were specific to M3GNet and did not occur 
with any of the other models tested.

The second motif comprises less common heteroaromatic rings 
outside metal coordination spheres.
After relaxation with M3GNet, thiophene rings were consistently 
distorted: the C--S bonds were elongated to lengths characteristic 
of single bonds, and the S atom was slightly displaced from the 
plane of the carbon atoms (\eg{} COD entry \codid{2012496}).
A similar elongation of the C--S bonds was observed in thiazole 
rings (\eg{} COD entry \codid{1518265}).
Unlike the distortions of the $\pi$ ligands, those of the thiophene 
and thiazole rings occurred with both M3GNet and CHGNet, but not 
with the PET models.
The closely analogous selenophene rings, in contrast, were mostly 
preserved by M3GNet and the PET models, but distorted by CHGNet 
(\eg{} COD entry \codid{1503809}).

\subsubsection{Analysis of problematic input structures}
\label{subsec:problematic-inputs}

Relaxation with an MLIP can help to detect structural errors that 
would otherwise be difficult to identify automatically at scale.
Once the model-specific failures described above are taken into 
account, substantial changes in the cell parameters, atomic positions 
or energy during relaxation can indicate deficiencies in the 
deposited structure model.
To evaluate how effectively the four models detect such deficiencies, 
we considered four types of error: missing hydrogen atoms, spurious 
hydrogen atoms, incorrectly assigned atom types, and incorrect 
atomic coordinates.
Entries with errors of the first three types had previously been 
identified during the manual curation of the COD based on chemical 
validation~\cite{Vaitkus2023}, whereas the entries with incorrect 
coordinates were found during the manual inspection of the M3GNet 
relaxations in this work.

Structures with missing hydrogen atoms lack some of the hydrogen 
atoms required by the expected coordination or protonation state.
During relaxation, the heavy atoms that would be bonded to the 
missing hydrogen atoms move substantially.
They either rearrange towards a geometry compatible with their 
reduced coordination (\eg{} an under-protonated \textit{sp}$^3$ 
carbon atom adopting a linear, \textit{sp}-like geometry; COD 
entry \codid{2100092}) or approach each other to form new chemical 
entities (\eg{} three under-protonated water molecules rearranging 
into an ozone-like unit; COD entry \codid{7224078}).
All four models produced substantial structural changes for 
entries of this type, which shows that relaxation-based 
screening can identify missing hydrogen atoms.

Spurious hydrogen atoms result from modelling errors in which 
additional hydrogen atoms are attached to heavy atoms that are 
already appropriately coordinated.
Because a MLIP operates on the given atomic configuration without 
knowledge of the intended composition, relaxation can reveal such 
an inconsistency through the displacement of the affected heavy 
atoms.
The magnitude of the displacement depends on the model and on the 
local environment of the spurious hydrogen atom.
A hydrogen atom placed between two bonded heavy atoms can push 
them apart beyond typical bonding distances (COD entry \codid{1519180}), 
whereas an additional hydrogen atom on a benzene ring may only 
slightly perturb the carbon atom to which it is attached, lengthening 
its two C--C bonds towards single-bond values while largely preserving 
the planarity of the ring (COD entry \codid{4061034}).
All four models produced detectable heavy-atom displacements for at 
least some of these entries, with the largest displacements for the 
PET models and the smallest for CHGNet.
Relaxation thus does not reliably reveal every spurious hydrogen atom, 
but it detects a subset of these errors.

An incorrectly assigned atom type is an atomic site assigned to an 
element other than the one present in the crystal.
Such errors usually replace an element by another with similar X-ray 
scattering power but not necessarily similar bonding preferences 
(\eg{} C$\leftrightarrow$N$\leftrightarrow$O).
The test set contained five such cases: an \textit{sp}$^2$-hybridised 
N atom modelled as C (\codid{4086432}), an \textit{sp}$^3$-hybridised 
B atom modelled as C (\codid{7117360}), an S atom in a bent geometry 
modelled as Cl (\codid{1519180}), and two cases in which the elements 
at two sites were interchanged, P$\leftrightarrow$S (\codid{7012760}) 
and P$\leftrightarrow$Cl (\codid{4065905}).
Relaxation of the structures with the N$\rightarrow$C and B$\rightarrow$C 
substitutions did not change the atomic positions substantially with 
any of the models.
In the remaining three cases, M3GNet performed best, substantially 
displacing the atoms around the mistyped sites in all three structures.
For the P$\leftrightarrow$S structure, however, this displacement 
occurred only with the ASE-based workflow and not with the LAMMPS-based 
workflow, which shows that the detection of subtle errors can depend on 
the optimizer and its settings.
The PET models produced similarly substantial changes in all cases 
except the P$\leftrightarrow$S structure, whereas CHGNet did so only 
for the P$\leftrightarrow$Cl structure.
In the P$\leftrightarrow$S structure, the interchanged P and S atoms 
occupy sites that are uncommon but chemically plausible.
The absence of a change with the PET models may therefore reflect 
their ability to describe a broader range of chemical environments 
than M3GNet, rather than a deficiency of these models.

Incorrect atomic coordinates place atoms at unphysical positions 
in the crystal.
Inspection of the $d_{\max}$ outliers in the M3GNet relaxations led 
to two previously unreported cases.
In both, the coordinates in the original publications are correct, 
so the errors were most likely introduced as typographical errors 
when the CIFs were prepared.
The first case is a misplaced water molecule coordinated to U in 
b\v{e}hounekite (\codid{9014173}).
All four models moved the water molecule towards its experimentally 
determined position; the RMSD exceeded 0.5~\AA{} in all relaxations,
 and $d_{\max}$ was approximately 1.45~\AA{} with M3GNet and exceeded 
 2~\AA{} with the other models.
The second case is a misplaced Cl atom in a thio-chloro-ruthenium 
cluster (\codid{2002167}).
Unlike the uranium compound, this error was detected only by some of 
the models: M3GNet and CHGNet moved the Cl atom by up to 1.9~\AA{} 
towards its experimentally determined position, whereas the PET models 
produced no substantial change in either RMSD or $d_{\max}$.
Both errors were reported to the COD maintainers and have since been 
corrected.

\begin{figure}[H]
    \includegraphics[width=\linewidth]{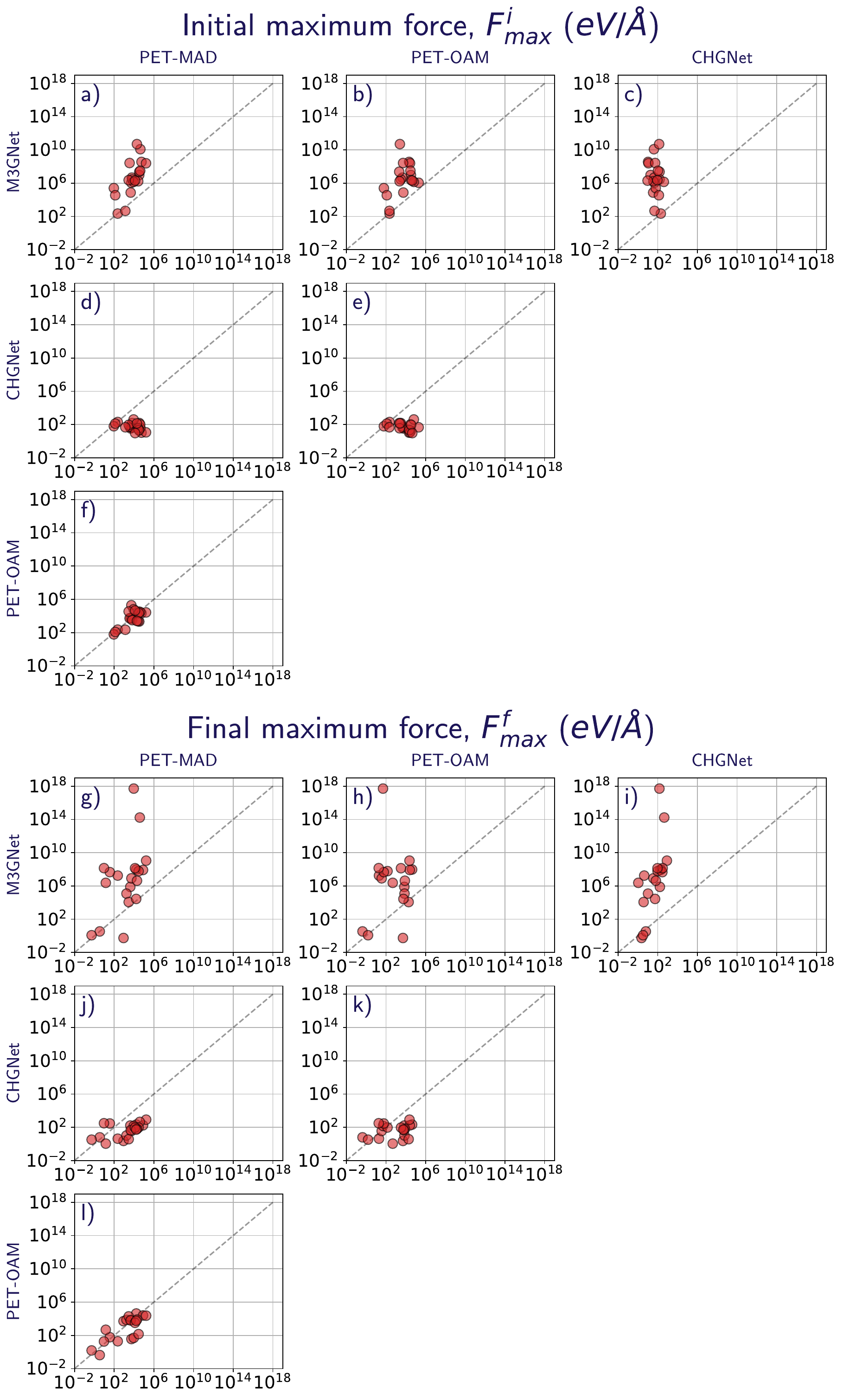}
    \caption{Initial and final largest atomic forces,
$F^{\mathrm{i}}_{\max}$ and $F^{\mathrm{f}}_{\max}$, for 
structures with duplicated atoms on special positions, 
compared pairwise between the four MLIPs.
The dashed line is the diagonal $y=x$.}
    \label{fig:Multi_Force_FormulaMismatch}
\end{figure}

During the relaxations, we also processed structures in which atoms 
on special positions were duplicated.
These structures contain too many atoms and are incorrect by 
construction, but they show how the models respond to atoms that 
are much too close to each other (Figure~\ref{fig:Multi_Force_FormulaMismatch}).
M3GNet predicts very large forces on such atoms, orders of magnitude 
larger than any other model.
CHGNet behaves differently from all other models: its $F_{\max}$ is 
nearly the same for all of these structures, even where the other 
models predict increasing forces.
The two PET models agree with each other and lie between these 
two extremes.

The energies of these structures show the same behaviour of CHGNet: 
all of them have nearly the same $E^{\mathrm{i}}$ and $E^{\mathrm{f}}$, 
regardless of the values obtained with the other models 
(Figure~\ref{sup-fig:Multi_Energy_FormulaMismatch}).
For $E^{\mathrm{i}}$, the two PET models mostly agree with each other, 
whereas the comparison of M3GNet with the PET models shows three 
distinct groups of structures: those for which only the M3GNet energy 
is high, those for which only the PET energies are high, and those for 
which all models agree.
For $E^{\mathrm{f}}$, the same three groups appear in all pairwise 
comparisons of M3GNet, PET-MAD and PET-OAM.
In addition, two structures not shown in the figure have final M3GNet 
energies of $-2\times10^{5}$ and $-2\times10^{9}$~eV/atom; these 
values are unphysical and indicate a breakdown of the model for 
strongly overlapping atoms.

\section{Conclusion}
\label{sec:Conclusion}

We applied universal machine learning interatomic potentials 
to the Crystallography Open Database (COD) with two aims: to 
assess how reliably these models describe experimentally determined 
crystal structures, and to determine whether the changes resulting from 
relaxation can be used to flag erroneous entries.
The relaxation results are intended to accompany the COD entries, so 
that users can see whether a structure reported in the literature is 
confirmed by relaxation with a machine learning interatomic potential (MLIP).

Of the 522\,086 COD entries, 317\,457 (60.8\%) passed the data quality 
filter and were relaxed with M3GNet, and 314\,153 of these (99.0\%) were 
relaxed successfully, i.e.\ converged without a large change in cell volume.
Most of the entries that were not relaxed are disordered (168\,910, or 
32.4\% of all entries); relaxing them requires first constructing ordered 
models, which was outside the scope of this work.
A further 25\,177 ordered entries were excluded because their declared and 
calculated chemical formulae differ or because they contain unmodelled solvent.

For most of the relaxed structures, the relaxed and initial cell volumes 
are nearly identical, showing that M3GNet reproduces the cell volumes of 
a chemically diverse set of experimental structures closely.
Convergence and volume change provided simple and effective criteria for 
identifying relaxations that do not confirm the initial structure.
The remaining descriptors -- the initial and final energies and maximum 
forces, RMSD and  the maximum displacement of a single atom -- carry 
information about the validity of the structures, but a manual inspection 
of 327 structures showed that no threshold on a single descriptor separates 
valid from invalid structures.
The labelled structures provide a starting point for training a classifier on 
several descriptors at once.
Because only a few hundred structures are labelled, the criteria used in 
this work are not calibrated against ground truth, and the fraction of 
erroneous entries that they miss cannot yet be estimated.

The comparison of four MLIPs on a subset of 524 structures showed that the 
models generally agree on the initial forces, particularly within the MatGL 
and PET model families.
CHGNet predicts lower forces for strongly anisotropic lattices and displaces 
the atoms considerably less than the other models, which is consistent with 
a softer potential energy surface.
Manual inspection of the relaxations revealed model-specific blind spots: 
M3GNet systematically distorts cyclopentadienyl and other $\pi$ ligands in 
metal complexes, and both MatGL models distort thiophene and thiazole rings, 
whereas the PET models preserve these motifs.
Such blind spots must be taken into account before a change upon relaxation 
is interpreted as evidence of an error in the deposited structure.
With this caveat, all models produced clear structural changes for 
structures with missing hydrogen atoms, and at least some of them detected 
spurious hydrogen atoms, several incorrectly assigned atom types, and two 
previously unreported coordinate errors, which have since been corrected 
in the COD.
No single model detected all types of error: the PET models, for example, 
did not detect the misplaced Cl atom that M3GNet and CHGNet detected.

\section{\label{section:data-availability}Data Availability}
The relaxed structures were deposited in the Theoretical Crystallography 
Open Database (TCOD)~\cite{Merkys2014,Merkys2017} and linked to the 
corresponding COD entries.
\section{\label{section:Acknowledgements}Acknowledgements}
This project has received funding under the grant agreement with the 
Research Council of Lithuania (LMTLT) (Project No. S-ITP-24-7).

\bibliography{Biblio}

\clearpage
\newpage
\section{Supplementary Information}
\label{sec:Supplementary}

\subsection{Full data distribution}
\label{subsec:full_checked}
\begin{figure}[!htb]
    \includegraphics[width=\linewidth]{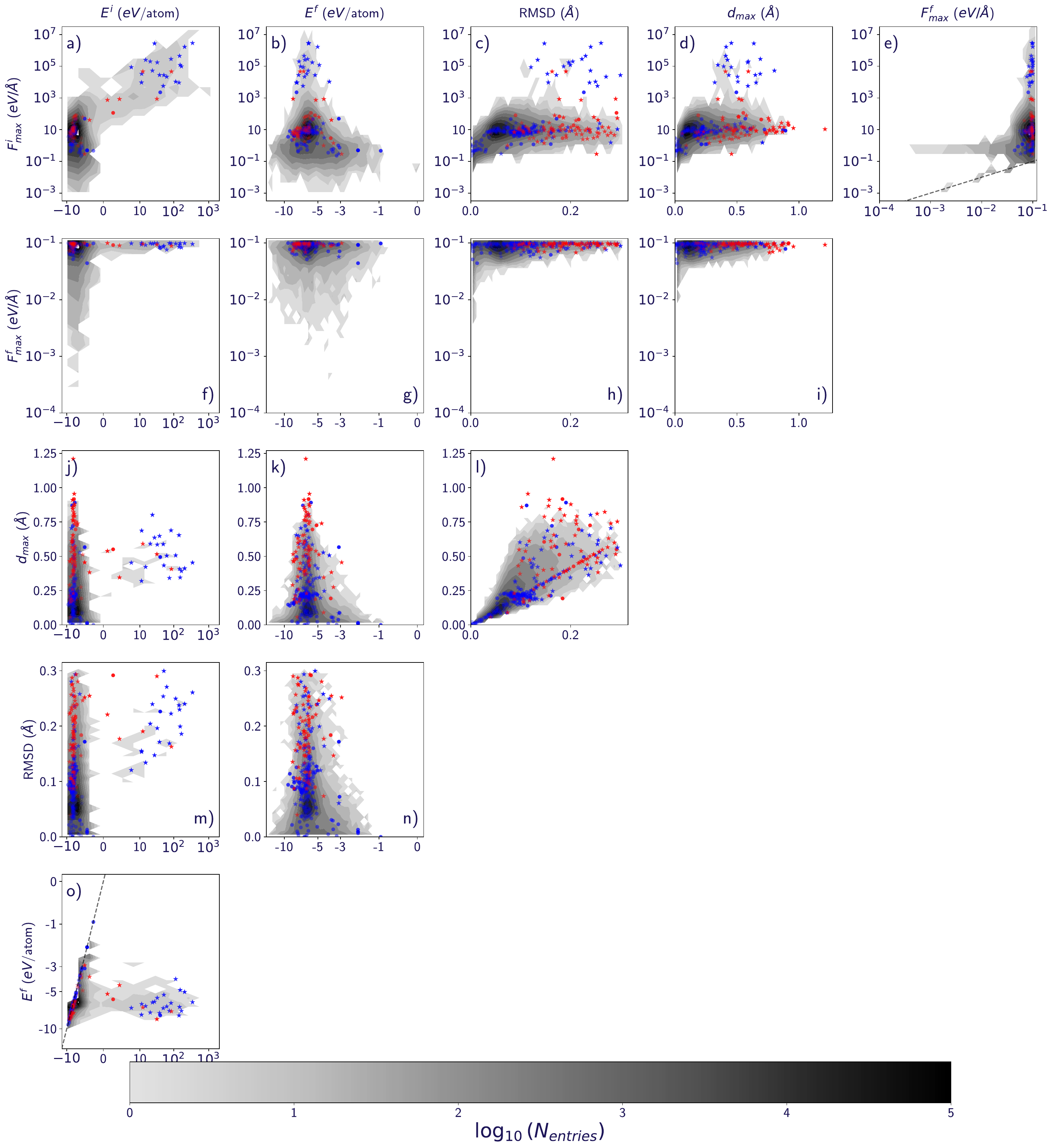}
    \caption{Pairwise distributions of the relaxation descriptors
            $E^{\mathrm{i}}$, $E^{\mathrm{f}}$, $F^{\mathrm{i}}_{\max}$, 
            $F^{\mathrm{f}}_{\max}$, RMSD and $d_{\max}$ for all 
            succesfully relaxed structures.
            Symbols mark the manually inspected structures: the colour 
            indicates the validity of the input structure (blue, valid; 
            red, invalid) and the shape the validity of the relaxed 
            structure (circle, valid; star, invalid).
            In the panels comparing initial with final values of $F_{\max}$ 
            and of $E$, the dashed line is the diagonal $y=x$. This plot shows 
            the wide range of possible values of energy and force encountered, 
            but the overwhelming majority of structures lie in the peaks.}
    \label{sup-fig:All_Graphs_full}
\end{figure}

Figure~\ref{sup-fig:All_Graphs_full} shows the full distribution of 
$E^{\mathrm{i}}$, $E^{\mathrm{f}}$, $F^{\mathrm{i}}_{\max}$, 
$F^{\mathrm{f}}_{\max}$, RMSD and $d_{\max}$.
As can be seen in the figure, $E^{\mathrm{i}}$ and 
$F^{\mathrm{i}}_{\max}$ have outliers that are very far away from 
the main peak populations.
Even though the outliers return to normal max force and energy at 
the end of relaxation, the manually checked structures in the outliers 
are all poor, with the exception of a single structure.

\subsection{Model pairwise comparison of Volume, Energies and Distances}
\label{subsec:multi_e_d}
\begin{figure}[htbp]
    \includegraphics[width=\linewidth]{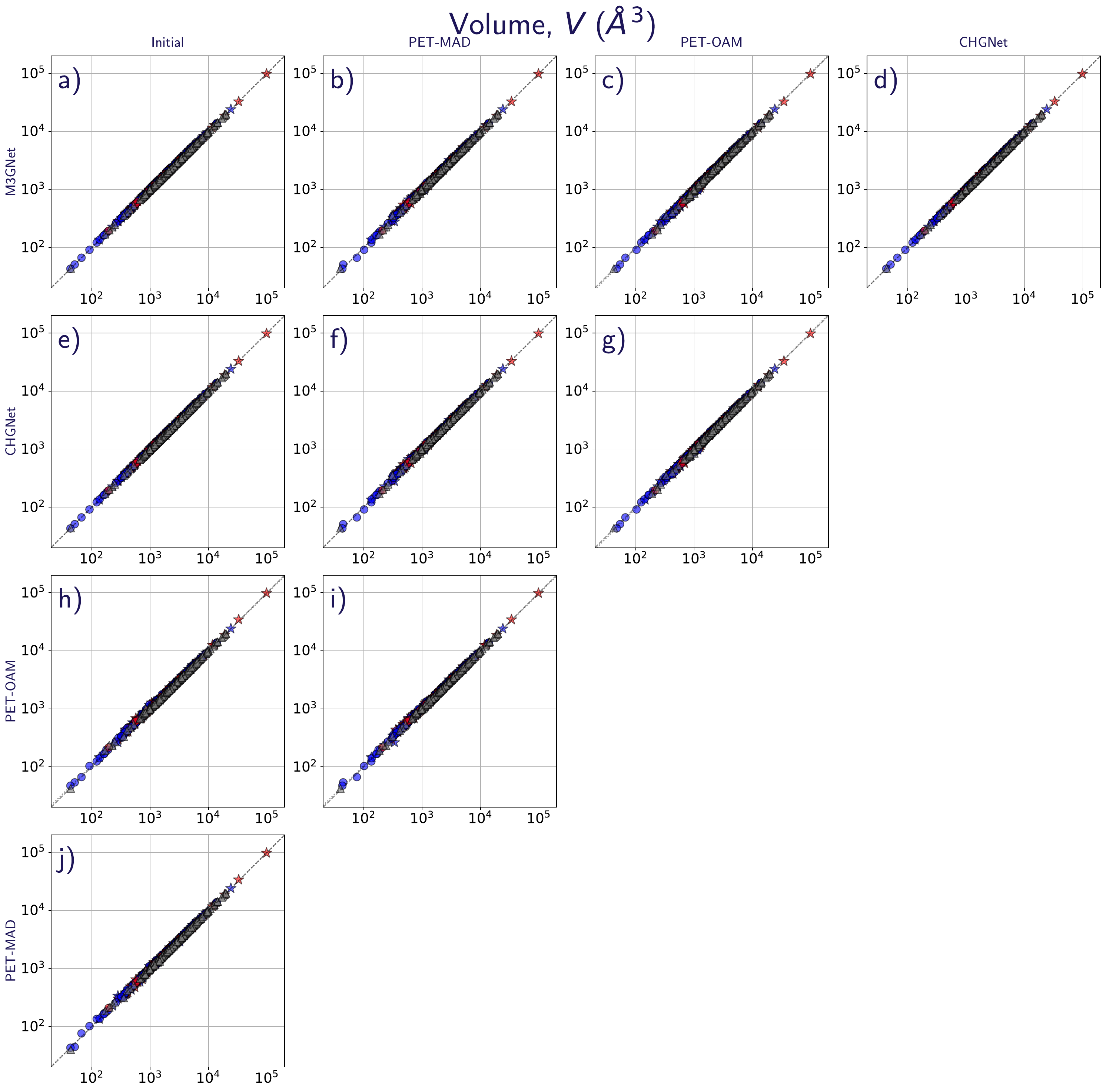}
    \caption{Pairwise correlation of relaxed cell volumes obtained with the 
    four MLIPs compared with the deposited volume.
Grey triangles denote structures that were not inspected manually.
For the inspected structures, the colour of the symbol indicates the validity 
of the input structure (blue, valid; red, invalid) and its shape the validity 
of the relaxed structure (circle, valid; star, invalid).
The dashed line is the diagonal $y=x$; the concordance correlation 
coefficients (CCC) are presented in the Supplementary Information 
Table~\ref{sup-table:multi_volume_stats}.}
    \label{sup-fig:Multi_Volume}
\end{figure}

\begin{figure}[!htb]
    \includegraphics[width=0.8\linewidth]{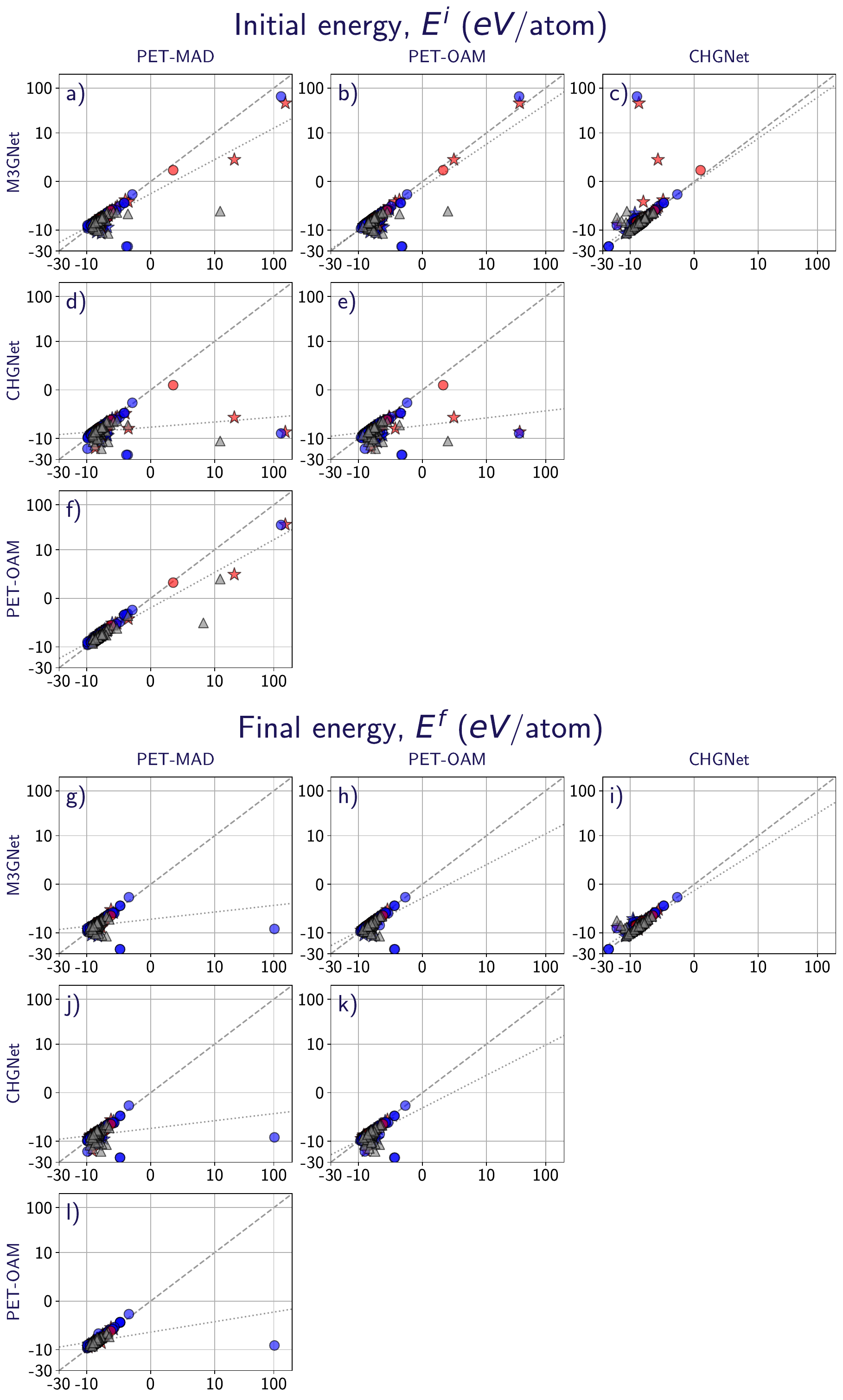}
    \caption{Pairwise correlation of initial and final energies,
            $E^{\mathrm{i}}$ and $E^{\mathrm{f}}$, obtained with the four MLIPs.
            Symbols are the same as in Figure~\ref{sup-fig:Multi_Volume}.
            The dashed line is the diagonal $y=x$ and the dotted line is the linear fit to $y=\operatorname{symlog}_{10}E$ (vertical axis) against $x=\operatorname{symlog}_{10}E$ (horizontal axis); the fitted equation and the coefficient of determination $R^2$ are presented in the Supplementary Information Table~\ref{sup-table:multi_energy_stats}.}
    \label{sup-fig:Multi_Energy}
\end{figure}

\begin{figure}[!htb]
    \includegraphics[width=\linewidth]{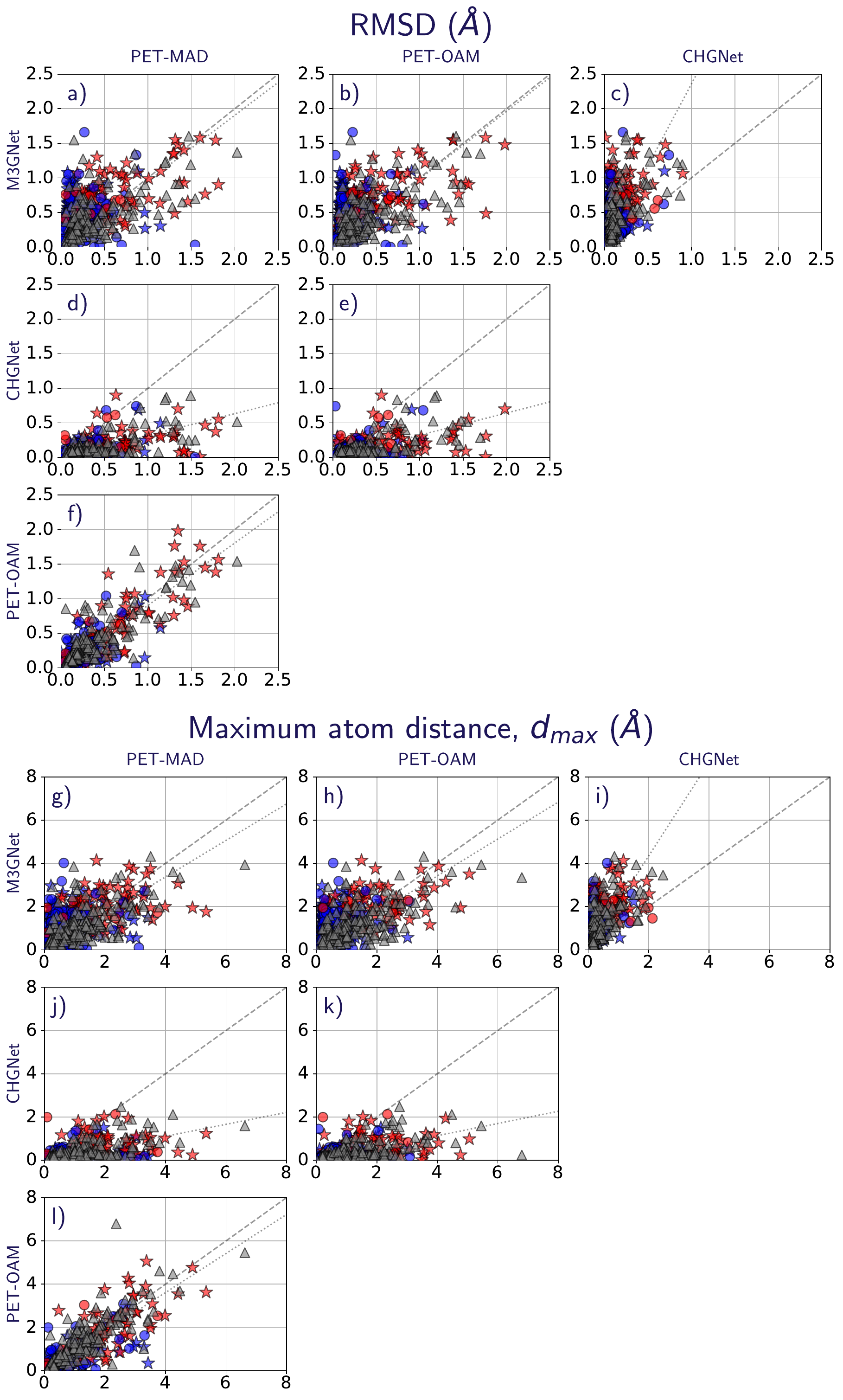}
    \caption{Pairwise correlation of RMSD and $d_{max}$, obtained with the four MLIPs.
    Symbols are the same as in Figure~\ref{sup-fig:Multi_Volume}.
    The dashed line is the diagonal $y=x$ and the dotted line is the 
    linear fit; the fitted equation and the coefficient of determination 
    $R^2$ are presented in the Supplementary Information Table~\ref{sup-table:multi_dist_stats}.}    
    \label{sup-fig:Multi_Distance}
\end{figure}

Figures~\ref{sup-fig:Multi_Volume},~\ref{sup-fig:Multi_Energy} 
and~\ref{sup-fig:Multi_Distance} show the full volume, energy and 
distance pairwise correlations for the 4 MLIPs tested. 
Table~\ref{sup-table:multi_volume_stats} contains the values of Lin’s 
concordance correlation coefficient for volume pairwise correlations and 
Tables~\ref{sup-table:multi_force_stats}, \ref{sup-table:multi_energy_stats} 
and \ref{sup-table:multi_dist_stats} contain the values of coefficient of 
determination ($R^2$) and the equation of linear fit.

As discussed in the results Section~\ref{subsec:crosscomparisons} the volumes 
are very strongly correlated, the energies mostly lie on the diagonal, and most 
of the atom displacement distances are in general agreement, apart from CHGNet. 
The other models agree on the sizes of RMSD and $d_{max}$, but the point to 
point agreement is very low. CHGNet shows a decrease in RMSD and $d_{max}$ by 
a factor of more than 2.

\begin{table}[htbp]
\centering
\caption{\label{sup-table:multi_volume_stats} Lin's concordance 
correlation coefficient (CCC) of the relaxed cell volumes, for each 
pair of models compared in Figure~\ref{sup-fig:Multi_Volume}.}
\begin{tabular*}{\linewidth}{@{\extracolsep{\fill}}ccc}
\hline
$x$ & $y$ & CCC\\
\hline
\hline
Initial & PET-MAD & 0.99954\\
\hline
Initial & PET-OAM & 0.99927\\
\hline
Initial & CHGNet & 0.99997\\
\hline
Initial & M3GNet & 0.99999\\
\hline
PET-MAD & PET-OAM & 0.99925\\
\hline
PET-MAD & CHGNet & 0.99955\\
\hline
PET-MAD & M3GNet & 0.99956\\
\hline
PET-OAM & CHGNet & 0.99931\\
\hline
PET-OAM & M3GNet & 0.99931\\
\hline
CHGNet & M3GNet & 0.99998\\
\hline
\end{tabular*}
\end{table}

\begin{table}[htbp]
\centering
\caption{\label{sup-table:multi_force_stats} Coefficient of 
determination ($R^2$) and linear fit of the maximum atomic force 
$\log_{10}(F_{max})$, for each pair of models compared in Figure~\ref{fig:Multi_Force}.}
\begin{tabular*}{\linewidth}{@{\extracolsep{\fill}}ccccc}
\hline
Quantity & $x$ & $y$ & $R^2$ & Equation \\
\hline
\hline
$F^i_{max}$ & PET-MAD & PET-OAM & 0.962 & $y=0.96x+0.07$ \\
\hline
$F^i_{max}$ & PET-MAD & CHGNet & 0.754 & $y=1.1x-0.31$ \\
\hline
$F^i_{max}$ & PET-MAD & M3GNet & 0.814 & $y=1.09x-0.25$ \\
\hline
$F^i_{max}$ & PET-OAM & CHGNet & 0.828 & $y=1.18x-0.43$ \\
\hline
$F^i_{max}$ & PET-OAM & M3GNet & 0.855 & $y=1.15x-0.33$ \\
\hline
$F^i_{max}$ & CHGNet & M3GNet & 0.939 & $y=0.93x+0.17$ \\
\hline
$F^f_{max}$ & PET-MAD & PET-OAM & 0.559 & $y=0.67x+0.17$ \\
\hline
$F^f_{max}$ & PET-MAD & CHGNet & 0.088 & $y=0.29x+0.25$ \\
\hline
$F^f_{max}$ & PET-MAD & M3GNet & 0.111 & $y=0.34x-0.24$ \\
\hline
$F^f_{max}$ & PET-OAM & CHGNet & 0.131 & $y=0.39x+0.18$ \\
\hline
$F^f_{max}$ & PET-OAM & M3GNet & 0.211 & $y=0.52x-0.32$ \\
\hline
$F^f_{max}$ & CHGNet & M3GNet & 0.613 & $y=0.82x-0.46$ \\
\hline
\end{tabular*}
\end{table}

\begin{table}[htbp]
\centering
\caption{\label{sup-table:multi_energy_stats} Coefficient of determination 
($R^2$) and linear fit of the per-atom energy $\operatorname{symlog}_{10}(E)$, 
for each pair of models compared in Figure~\ref{sup-fig:Multi_Energy}.}
\begin{tabular*}{\linewidth}{@{\extracolsep{\fill}}ccccc}
\hline
Quantity & $x$ & $y$ & $R^2$ & Equation \\
\hline
\hline
$E^i$ & PET-MAD & PET-OAM & 0.920 & $y=0.73x-0.2$ \\
\hline
$E^i$ & PET-MAD & CHGNet & 0.067 & $y=0.11x-0.8$ \\
\hline
$E^i$ & PET-MAD & M3GNet & 0.763 & $y=0.7x-0.26$ \\
\hline
$E^i$ & PET-OAM & CHGNet & 0.089 & $y=0.16x-0.76$ \\
\hline
$E^i$ & PET-OAM & M3GNet & 0.768 & $y=0.89x-0.13$ \\
\hline
$E^i$ & CHGNet & M3GNet & 0.221 & $y=0.92x-0.03$ \\
\hline
$E^f$ & PET-MAD & PET-OAM & 0.185 & $y=0.21x-0.67$ \\
\hline
$E^f$ & PET-MAD & CHGNet & 0.067 & $y=0.16x-0.76$ \\
\hline
$E^f$ & PET-MAD & M3GNet & 0.067 & $y=0.15x-0.75$ \\
\hline
$E^f$ & PET-OAM & CHGNet & 0.309 & $y=0.67x-0.33$ \\
\hline
$E^f$ & PET-OAM & M3GNet & 0.359 & $y=0.69x-0.29$ \\
\hline
$E^f$ & CHGNet & M3GNet & 0.761 & $y=0.82x-0.13$ \\
\hline
\end{tabular*}
\end{table}

\begin{table}[htbp]
\centering
\caption{\label{sup-table:multi_dist_stats} Coefficient of determination ($R^2$) and 
linear fit of RMSD and the maximum atomic displacement $d_{max}$, for each pair of models 
compared in Figure~\ref{sup-fig:Multi_Distance}.}
\begin{tabular*}{\linewidth}{@{\extracolsep{\fill}}ccccc}
\hline
Quantity & $x$ & $y$ & $R^2$ & Equation \\
\hline
\hline
RMSD & PET-MAD & PET-OAM & 0.849 & $y=0.9x$ \\
\hline
RMSD & PET-MAD & CHGNet & 0.630 & $y=0.32x$ \\
\hline
RMSD & PET-MAD & M3GNet & 0.612 & $y=0.95x$ \\
\hline
RMSD & PET-OAM & CHGNet & 0.611 & $y=0.32x$ \\
\hline
RMSD & PET-OAM & M3GNet & 0.610 & $y=0.98x$ \\
\hline
RMSD & CHGNet & M3GNet & 0.596 & $y=2.36x$ \\
\hline
$d_{max}$ & PET-MAD & PET-OAM & 0.843 & $y=0.9x$ \\
\hline
$d_{max}$ & PET-MAD & CHGNet & 0.562 & $y=0.28x$ \\
\hline
$d_{max}$ & PET-MAD & M3GNet & 0.662 & $y=0.84x$ \\
\hline
$d_{max}$ & PET-OAM & CHGNet & 0.562 & $y=0.28x$ \\
\hline
$d_{max}$ & PET-OAM & M3GNet & 0.649 & $y=0.85x$ \\
\hline
$d_{max}$ & CHGNet & M3GNet & 0.596 & $y=2.17x$ \\
\hline
\end{tabular*}
\end{table}

\subsection{Energies of structures with duplicated atoms}
\label{subsec:mismatch_energies}
\begin{figure}[!htb]
    \includegraphics[width=\linewidth]{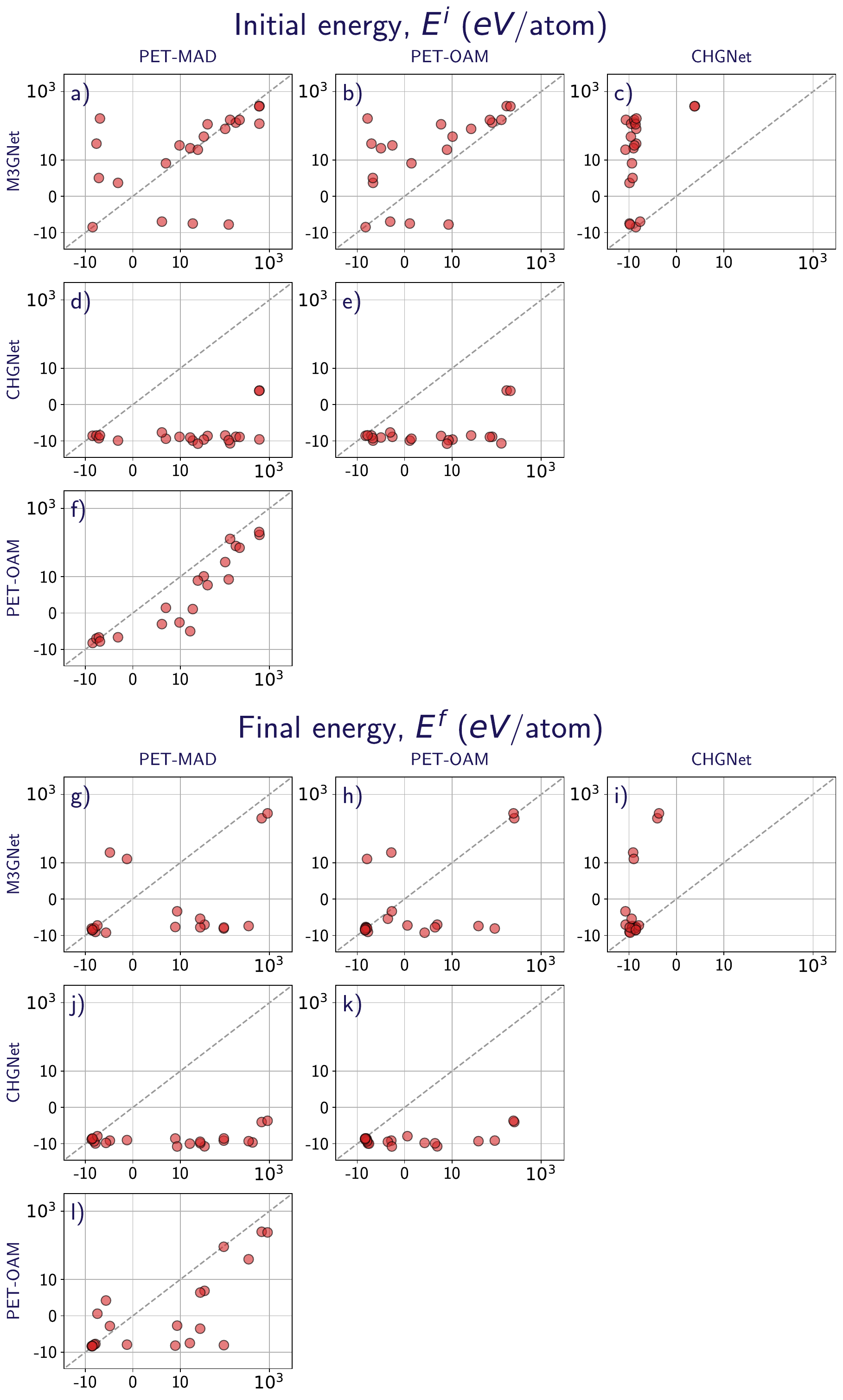}
    \caption{Initial and final energies,
            $E^{\mathrm{i}}$ and $E^{\mathrm{f}}$, for structures 
            with duplicated atoms on special positions, compared pairwise 
            between the four MLIPs.
            The dashed line is the diagonal $y=x$. Two structures for M3GNet 
            are not shown, as they have unphysical energies of 
            $E^{\mathrm{f}}=-2\times10^5$~eV/atom and $E^{\mathrm{f}}=-2\times10^9$~eV/atom.}
    \label{sup-fig:Multi_Energy_FormulaMismatch}
\end{figure}

Figure~\ref{sup-fig:Multi_Energy_FormulaMismatch} shows the pairwise 
correlation of $E^{\mathrm{i}}$ and $E^{\mathrm{f}}$.
The figure was discussed in Section~\ref{subsec:problematic-inputs} and 
shows a wide array of different behaviours from the models.
There is very little agreement between the models on the energies of these 
unphysical structures, with the closest agreement being found between 
PET-MAD and PET-OAM for $E^{\mathrm{i}}$.

\end{document}